\documentclass[prl, twocolumn]{revtex4}
\usepackage{graphicx}% Include figure files
\usepackage{dcolumn}% Align table columns on decimal point
\usepackage{bm}% bold math
\input{epsf}

\long\def\comment#1{}

\begin{document}
\title{Adaptive Boolean Networks and Minority Games with Time--Dependent Capacities}
\author{Aram~Galstyan and  Kristina~Lerman}
\affiliation{
Information Sciences Institute\\
University of Southern California\\
4676 Admiralty Way, Marina del Rey, CA 90292\\
}

%\date{March 1, 2001}   % Deleting this command produces today's date.

\begin{abstract}
In this paper we consider a network of boolean agents that compete
for a limited resource. The agents play the so called Generalized
Minority Game where the capacity level  is allowed to vary
externally. We study the properties of such a system for different
values of the mean connectivity $K$ of the network, and show that
the system with $K=2$ shows a high degree of coordination for
relatively large variations of the capacity level.
\end{abstract}

\maketitle

Complex adaptive systems composed of agents under mutual influence
have attracted considerable interest in recent years. A few
examples that have been  studied extensively are genetic
regulatory networks~\cite{Kauffman}, ecosystems~\cite{Levin}, and
financial markets~\cite{Anderson}. These kinds of systems
often display rich and complex dynamics and have been shown
to posses global properties that cannot be simply deduced from the
details of the microscopic behavior of individual agents.

The Minority Game~\cite{Challet1997} (MG) is one of the simplest
examples of a complex dynamical system. It was introduced by
Challet and Zhang as a simplification of Arthur's El Farol Bar
attendance problem~\cite{Arthur1994}. The MG consists of $N$
agents with bounded rationality that repeatedly choose between two
alternatives labelled $0$ and $1$ ({\em e.g.}, staying at home or
going to the bar). At each time step, agents who made the minority
decision win. In the Generalized Minority Game~\cite{Johnson}, the
wining group is $1$ ($0$) if the fraction of the agents who chose
``1'' is smaller (greater) than the capacity level $\eta$,
$0<\eta<1$ (for $\eta=0.5$, the game reduces to the the
traditional MG). Each agent uses a set of $S$ strategies to decide
its next move and reinforces strategies that would have predicted
the winning group. A strategy is simply a lookup table that
prescribes a binary output for all possible inputs. In the
original version of the game, the input is a binary string
containing the last $m$ outcomes of the game, so the agents
interact by sharing the same global signal. If the agents choose
either action with probability $1/2$ (the random choice game),
then, on average, the number of agents choosing ``1'' (henceforth
referred to as attendance) is $(N-1)/2$ with standard deviation
$\sigma=\sqrt{N}/2$ in the limit of large $N$. The most
interesting phenomenon of the minority model is the emergence of a
coordinated phase, where the standard deviation of attendance, the
volatility, becomes smaller than in the random choice game. The
coordination is achieved for  memory sizes for which  the
dimension of the reduced strategy space is comparable to the
number of agents in the system\cite{Challet1998,Savit1999}, $2^m
\sim N$. It was later pointed out \cite{Cavagna1998} that the
dynamics of the game remains mostly unchanged if one replaces the
string with the actual histories with a random one, provided that
all the agents act on the same signal. Analytical studies based on
this simplification have revealed many interesting properties of
the minority model\cite{Challet1999,MinorityGameURL}.

In addition to the original MG, different versions of the game
where the agents interact using local information only (cellular
automata~\cite{Kalinowski2000}, evolving random boolean
networks~\cite{Paczuski}, personal histories~\cite{deCara}), have
been studied. In particular, it was established that coordination
still arises out of local interactions, and the system as a whole
achieves ``better than random'' performance in terms of the
utilization of resources. Note that although the minority game was
introduced as a toy model of the financial markets, it can serve
as a general paradigm for resource allocation and load balancing
in multi--agent systems.

In all previous studies the capacity level has been fixed as an
external parameter, so the environment in which the agents compete
is stationary. In many situations, however, agents have to operate
in dynamic (and in general, stochastic) environments. It is
interesting to see if  coordinated behavior still emerges, and to
what degree agents  can adapt to the changing environment. We
address this problem in the present paper. Namely, we study a
system of boolean agents playing a generalized minority game,
assuming that the capacity level is not fixed but varies with
time, $\eta(t)=\eta_0+\eta_1(t)$, where $\eta_1(t)$ is a time
dependent perturbation. The framework of the interactions is based
on  Kauffman NK random boolean nets~\cite{Kauffman}, where each
agent gets its input from $K$ other randomly chosen agents, and
maps the input to a new state according to a boolean function of
$K$ variables, which is also randomly chosen and quenched
throughout the dynamics of the system. The generalization we make
is that agents are allowed to adapt by having more than one
boolean function, or strategy, and the use of a particular
strategy is determined by an agent based on how often it predicted
the winning group throughout the game. Note that this approach is
very different from adaptation through evolution studied
previously in the context of the minority model~\cite{Paczuski}.

Our main observation is that networks with small $K$ ($K<5$) adapt
to a certain degree to the changes in the capacity level. In
particular, networks with $K=2$ show a tendency towards
self--organization into a phase characterized by small
fluctuations, hence, an efficient utilization of the resource,
even for relatively large variations in the capacity level
$\eta(t)$. Note, that in the Kauffman nets with $K>2$ the dynamics
of the system is chaotic with an exponentially increasing length
of attractors as the system size grows, while for $K<2$ the
network reaches a frozen configuration. The case $K=2$ corresponds
to a phase transition in the dynamical properties of the network
and is often referred as the ``edge of the chaos''\cite{Kauffman}.
We would like to reiterate, however, that our system is different
from a Kauffman network since the agents have an internal degree
of freedom, characterized by their strategies. Specifically, our
system does not necessarily have periodic attractors, while in
Kauffman nets periodic attractors are guaranteed to exist due to
the finite phase space and quenched rules of updating.

Let us consider a set of $N$ boolean agents described by
 ``spin'' variables $s_i=\{0,1\}$, $i=1,\ldots,N$. Each agent gets
 its input from $K$ other randomly chosen agents, and maps the
 input to a new state:
\begin{equation}
s_i(t+1)=F_i^j(s_{k_1}(t), s_{k_2}(t), ...,s_{k_K}(t))
\end{equation}
where $s_{k_i}$, $i=1,\ldots,K$ are the set of neighbors, and
$F_i^j, j=1,\ldots,S$ are randomly chosen  boolean functions
(called strategies hereafter) used by the $i$-th agent. For each
strategy $F_i^j$, the agent keeps a score that monitors the
performance of that strategy, adding (subtracting) a point if the
strategy predicted the winning (loosing) side. Let the
``attendance'' $A(t)$ be the cumulative output of the system at
time $t$, $A(t)=\sum_{i=1}^{N}s_i(t)$. Then the winning choice is
$"1"$ if $A(t)\le N\eta(t)$, and $"0"$ otherwise. Those in the
wining group are awarded a point while the others loose one.
Agents play the strategies that have predicted the winning side
most often, and the ties are broken randomly.

As a global measure of optimality  we consider
$\delta(t)=A(t)-N\eta(t)$, that describes the deviation from the
optimal resource utilization. We are primarily interested in the
cumulative ``waste'' over a certain time window:
\begin{equation}
\sigma=\sqrt{\frac{1}{T}\sum_{t=t_0}^{t_0+T}\delta(t)^2}
\end{equation}
For $\eta_1(t)=0$ this quantity is simply the volatility as
defined in the traditional minority game. We compare the
performance of our system to a default random choice game, defined
as follows: assume that the agents are told what is the capacity
$\eta(t)$ at time $t$, and they choose to go to the bar with
probability $\eta(t)$. In this case the main attendance will be
close to $\eta(t)N$ at each time step, and the fluctuations around
the mean are given by the standard deviation
\begin{equation}
\sigma_0^2=N\frac{1}{T}\int_{t_0}^{t_0+T}dt^{\prime}\eta(t^{\prime})[1-\eta(t^{\prime})]
\label{sigma0}
\end{equation}

We performed intensive numerical simulations of the system
described above, with the number of agents ranging from $100$ to
$10^4$, and for network connectivity $K$ ranging from $1$ to $10$.
Although in our simulations we used different forms for the
perturbation $\eta_1(t)$, in this paper we consider periodic
perturbations only. For each $K$, a set of strategies was chosen
for each agent randomly and independently from a pool of $2^{2^K}$
possible boolean functions, and was quenched throughout the game.
In all simulations we used $S=2$ strategies per agent. Starting
from a random initial configuration, the system evolved according
to the specified rules. The duration of the simulation $T_0$ was
determined by the particular choice of $\eta(t)$. Depending on the
amplitude of the perturbation, we run the simulations for $10$ to
$20$ periods, and usually used the data for the last two periods
to determine $\sigma$.

\begin{figure}[tp]
\epsfxsize = 2.50in \center{\epsffile{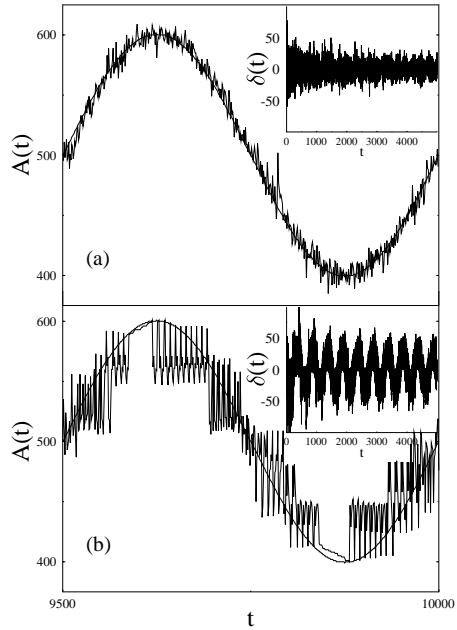}} \caption{A segment
of the attendance  time series for $\eta (t)=0.5+0.1sin(2 \pi
t/T)$,$T=500$; a) Boolean network with $K=2$, b) traditional
(generalized) minority game with $m=6$. The insets show the
respective time series of the deviation $\delta(t)$.} \label{fig1}
\end{figure}

Figure~\ref{fig1} (a) shows a typical segment of the time series
of the attendance $A(t)$ for a system of size $N=1000$, network
connectivity $K=2$, and a sinusoidal perturbation $\eta_1(t)$. One
can see that the system is efficient --- it adapts very quickly to
changes in the capacity level. The inset shows the time series of
the deviation $\delta(t)$. Initially there are strong
fluctuations, hence poor utilization of the resource, but after
some transient time the system as a whole adapts and the strength
of the fluctuations decreases. In particular, for the system sizes
considered in this paper (up to $N=10^4$) $\sigma$ is considerably
smaller than the standard deviation $\sigma_0$ in the random
choice game. This should not hold for sufficiently large $N$,
however, since $\sigma/\sqrt{N}$ increases slowly with $N$ (see
below), while for the random choice game
$\sigma_0\propto\sqrt{N}$. Note also, that the agents have
information only about the winning choice, but not the capacity
level. This suggests that the particular form of the perturbation
may not be important as long as it meets some general criteria for
smoothness.

We also studied the effect of the changing capacity level in the
traditional (generalized) minority model with publicly available
information about the last $m$ outcomes of the game. We plot the
attendance and deviation time series for a system with a memory
size $m=6$ (corresponding to the minimum of $\sigma$) in
Fig.~\ref{fig1} (b). One can see that in this case the system
also reacts to the external change; however, the overall performance in terms
of efficiency of resource allocation, as described by $\sigma$, is much poorer
compared to the previous case.

\begin{figure}[bp]
\epsfxsize = 2.50in \center{\epsffile{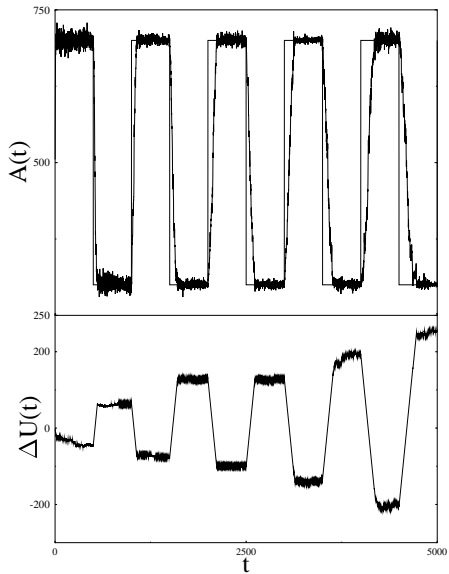}} \caption{Time series of attendance (top) and the gap
in strategy scores (bottom) for the square-shaped capacity level variations}
\label{fig2}
\end{figure}
 Another interesting observation is that if we run the simulations long enough,
 the response of the system to the changing capacity level gets ``out of phase'' with the
perturbation, leading to a gradual deterioration in the
performance of the system, and the time during which the efficient
phase is stable strongly depends on the rate of the changes in the
capacity level, as well as on the number of agents in the system.
Our results suggest that this effect is due to the increasing gap
in strategy scores. As the gap in strategy scores grows, it
becomes increasingly difficult for an agent to abandon a
previously more successful strategy that has stopped performing
well as the capacity level changes, because it takes longer for a
previously loosing strategy to accumulate enough points to be
played. We demonstrate this effect on a simpler, square-like
perturbation depicted in Fig.~2. One can see that each time the
capacity ``jumps'' to its new value, it takes longer for the
agents to adapt to the change. To illustrate why this happens, we
plot the evolution of the gap (difference) $\Delta U(t)$ in
strategy scores for one agent. For pedagogical reasons, we chose
an agent with the simplest anti-correlated strategies: one of
whose strategies always chooses ``0'' and the other ``1'',
regardless of input. As the amplitude of the oscillations in score
difference grows in time, it takes longer for the agent to switch
between strategies. The same is true for the difference between
strategy scores averaged over all agents, resulting in a growing
lag between attendance and the new capacity level after a
``jump''. Remarkably, one can get rid of the dephasing effect
simply by introducing an upper and lower bounds for the strategy
scores, thus, limiting their maximum difference.

\begin{figure}[hptb]
\epsfxsize = 3.0in \center{\epsffile{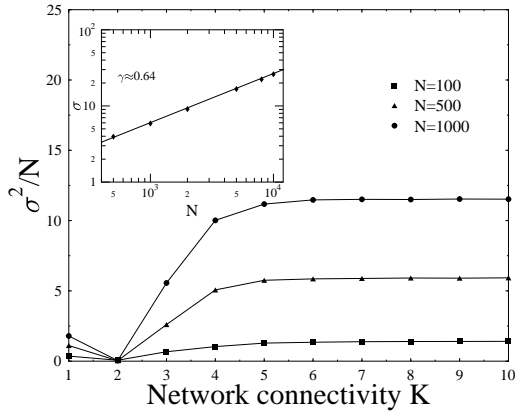}}
\caption{$\sigma^2/N$ vs the network connectivity for different
system sizes and $\eta (t)=0.5+0.15sin(2 \pi t/T)$,$T=1000$. Inset
plot shows the scaling relationship between $\sigma$ and $N$ for
 $K=2$. Average over 16 runs has been taken.}\label{fig3}
\end{figure}

In Fig.~\ref{fig3} we plot the variance per agent versus network
connectivity $K$, for system sizes $N=100, 500, 1000$. For each
$K$ we performed $32$ runs and averaged results. Our simulations
suggest that the details of this dependence are not very sensitive
to the particular form of the perturbation $\eta_1(t)$, and the
general picture is the same for a wide range of functions,
provided that they are smooth enough. As we already mentioned, the
variance attains its minimum for $K=2$, independent of the number
of agents in the system.  For bigger $K$ it saturates at a value
that depends on the amplitude of the perturbation and  on the
number of agents in the system. We found that for large $K$ the
time series of the attendance closely resembles the time series in
the absence of perturbation. This implies that for large $K$ the
agents do not ``feel'' the change in the capacity level.
Consequently, the standard deviation increases linearly with the
number of agents in the system, $\sigma\propto N$. For $K=2$, on
the other hand, $\sigma$ increases considerably slower with the
number of agents in the system, $\sigma\propto N^{\gamma}$,
$\gamma <1$ (see the inset in Fig. 3). Our results indicate that
the scaling (i.e., the exponent $\gamma$) is not universal and
depends on the perturbation.

\begin{figure}[hptb]
\epsfxsize = 3.0in \center{\epsffile{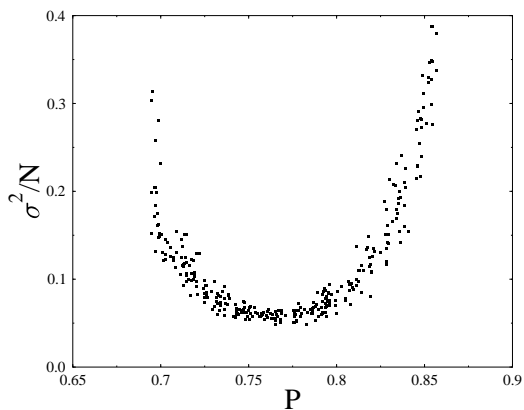}} \caption{Standard
deviation per agent vs P for K=3 networks: N=1000,
$\eta(t)=0.5+0.15\sin(2 \pi t/T)$, $T=1000$} \label{fig4}
\end{figure}

Though the results presented here look very interesting, we
currently do not have an analytical theory for the observed
emergent coordination. In contrast to the traditional minority
game, where global interactions  and the Markovian approximation
allow one to construct a mean field description, our model seems
to be analytically intractable due to the explicit emphasis on
local information processing. We strongly believe, however, that
the adaptability of the networks with $K=2$ is related to the
peculiar properties of the corresponding Kauffman nets, and
particularly, to the phase transition between the chaotic and
frozen phases. It is known \cite{Derrida} that the phase
transition in the Kauffman networks can be achieved by tuning the
homogeneity parameter $P$ which is the fraction of 1's or 0's  in
the output of the boolean functions (whichever is greater), with
the critical value given by $P_c=1/2+1/2\sqrt{1-2/K}$. To test our
hypothesis, we studied the properties of networks with $K=3$ for a
range of homogeneity parameter $P$. In Fig.~\ref{fig4} we plot
$\sigma^2/n$ versus the homogeneity parameter $P$. One can see
that the optimal resource allocation is indeed achieved in the
vicinity of the $P_c\approx 0.78$.

In conclusion, we studied a network of adaptive boolean agents
competing in a dynamic environment. We established that networks
with connectivity $K=2$ can be extremely adaptable and robust with
respect to capacity level changes. For $K>2$ the coordination can
be achieved by tuning the homogeneity parameter to its critical
value. Remarkably, adaptation happens without the agents
knowing the capacity level. Interestingly, the system that uses
local information is much more efficient in a dynamic environment
than a system that uses global information. This suggests that our
model may serve as a feasible mechanism for distributed resource
allocation in multi-agent systems.

{\bf Acknowledgements}\\
The authors would like to thank D. Tsigankov,  J. Crutchfield and
C. Shalizi for their helpful suggestions. The research reported
here was supported by the Defense Advanced Research Projects
Agency (DARPA) under contracts number F30602-00-2-0573, and in
part by the National Science Foundation under Grant No. 0074790.

\end{document}